\documentclass[preprint]{aastex63}
\usepackage{graphicx}
\usepackage{comment}
\usepackage{sidecap}
\usepackage{epstopdf}
\graphicspath{{figures/}} 

\newcommand{\rstar}{\ensuremath{R_\star}}

\newcommand{\rhostar}{\ensuremath{\rho_\star}}

\newcommand{\rpl}{\ensuremath{R_{\rm p}}}

\newcommand{\rearth}{\ensuremath{R_\earth}}

\newcommand{\Kepler}{\textit{Kepler}}

\newcommand{\K}{\ensuremath{K}}

\received{}
\revised{}
\accepted{}
\submitjournal{ApJL}
\shorttitle{The Kepler Peas in a Pod Pattern is Astrophysical}
\shortauthors{Weiss \& Petigura}

\begin{document}

\title{The Kepler Peas in a Pod Pattern is Astrophysical}

\correspondingauthor{Lauren M. Weiss}
\email{weisslm@hawaii.edu}

\author[0000-0002-3725-3058]{Lauren M. Weiss}
\affiliation{Institute for Astronomy, 2680 Woodlawn Dr., Honolulu, HI 96822, USA}
\author[0000-0003-0967-2893]{Erik A. Petigura}
\affil{Department of Physics \& Astronomy, University of California Los Angeles, Los Angeles, CA 90095, USA}

\begin{abstract}
\Kepler\ planets around a given star have similar sizes to each other and regular orbital spacing, like ``peas in a pod.''  Several studies have tested whether detection bias could produce this apparent pattern by resampling planet radii at random and applying a sensitivity function analogous to that of the Kepler spacecraft.  However, \citet{Zhu2019} argues that this pattern is not astrophysical but an artifact of Kepler's discovery efficiency at the detection threshold.  To support this claim, their new analysis samples the transit signal-to-noise ratio (SNR) to derive a synthetic population of bootstrapped planet radii. Here, we examine the procedure of sampling transit SNR and demonstrate it is not applicable. Sampling transit SNR does not set up random, independent planet radii, and so it is unsuitable for corroborating (or falsifying) detection bias as the origin of apparent patterns in planet radius.  By sampling the planet radii directly and using a simple model for Kepler's sensitivity, we rule out detection bias as the source of the peas-in-a-pod pattern with $>10$-$\sigma$ confidence.

\end{abstract}


\section{Introduction} \label{sec:intro}
The NASA \Kepler\ Mission detected hundreds of small planets within 1 au in multi-planet systems \citep{Borucki2011,Lissauer2011_multis,Rowe2014}.  The patterns in these multi-planet system architectures, or lack thereof, provide key information about the assembly and subsequent evolution of small planets close to their stars.

In \Kepler's multi-planet systems, the size of a transiting planet is correlated with the size of its detected transiting neighbors, as first reported in \citet{Lissauer2011_multis}. The improved accuracy in 909 planetary parameters and 355 stellar parameters obtained through the California \Kepler\ Survey \citep[CKS]{Petigura2017, Johnson2017} enabled a more detailed examination of this pattern \citep[][W18]{Weiss2018}. \citet{Millholland2017}\footnote{This paper was written after W18, but accepted for publication earlier.} performed a complementary analysis in systems for which transit timing variations have been detected and found that, in those systems, both the planet radius and mass are correlated with the radius and mass of the nearest neighboring planet.

\added{Although self-similarity of planet sizes is common, W18 found that in 63\% of adjacent pairs, the outer planet is larger than the inner planet.  This result is consistent with earlier work: \citet{Ciardi2013} found that in 68\% of pairs (either adjacent or non-adjacent) with at least one planet larger than 3\,\rearth, the outer planet is larger than the inner planet.  Conversely, there is no preferred ordering of planet sizes in pairs where both planets are smaller than 2\,\rearth.  Note that size similarity and larger outer planets can be present in the same system, and even in the same pair of planets.}

\begin{SCfigure}
    \caption{Systems from the California-Kepler Survey with 4 or more transiting planets.  Each row corresponds to a planetary system, with the star KOI number at the left, and planets represented with their measured semi-major axis (x-axis) and physical radius (point size).  The color corresponds to equilibrium temperature.  The systems are ranked by stellar mass, for which the errors were typically 5\%.  In many systems, the planets are similar in size to their neighbors and have regular orbital spacing.  Reproduced from W18.}
    \includegraphics[height=\textheight]{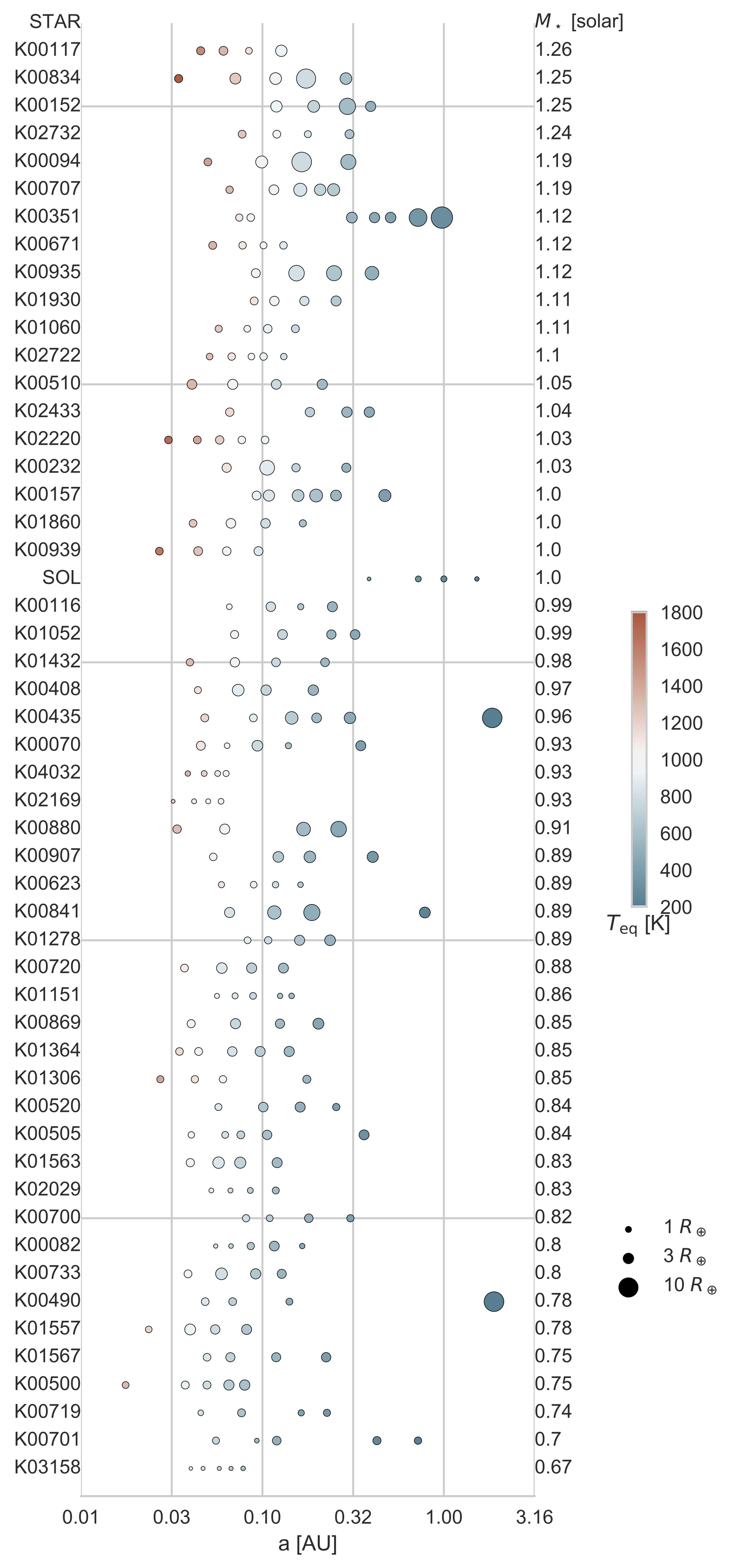}
    \label{fig:peas}
\end{SCfigure}

In \Kepler's multi-planet systems, the orbital spacing between transiting planets is regular (W18), but with no preference for mean motion resonances \citep{Fabrycky2014}.  Also, the smallest planets tend to have the closest orbital spacings, although the combination of dynamical stability and detection biases are not sufficient to explain the spacings of the detected planets (W18).  Our shorthand way of describing these patterns---the self-similar planet sizes, the self-similar period ratios of planets, and the relationship between planet size and spacing---in combination is that the \Kepler\ multi-planet systems resemble ``peas in a pod.''  Figure \ref{fig:peas} shows the high-multiplicity systems, where the pattern is apparent in many of the systems.

If this peas-in-a-pod pattern is based on the underlying distribution of planet sizes, then the pattern is a clear signature of the physical processes that govern the assembly of planetary systems.  However, a significant concern is whether the correlation between a planet's size and the sizes of its neighbors could result from detection bias.  In W18, we examined the role of detection bias by conducting a series of null-hypothesis tests via bootstrap resampling.  We found that detection bias could not explain the observed patterns and concluded that the patterns are indeed astrophysical.   In addition, full forward-modeling studies, which simultaneously model many attributes of the proposed underlying planetary systems' architectures, have found that models in which planets have similar sizes to their neighbors and regular orbital spacing better reproduce the \Kepler\ data than models in which the planet sizes or spacing are random \citep{Mulders2018, Sandford2019, He2019}.  However, a recent manuscript by \citet[Z20]{Zhu2019} arrives at the conclusion that detection bias is the main (or perhaps only) source of the apparently correlated planet sizes and spacings.

Here, we examine the method of Z20 and test its applicability to the peas-in-a-pod pattern.  In \S\ref{sec:evidence} we present examples of peas-in-a-pod systems that cannot be explained by detection bias.  In \S\ref{sec:mock} we  examine the Z20 null-hypothesis testing method.  We find that one of the necessary assumptions for the Z20 analysis to work, the ability to sample random planet radii, was not met, and that their analysis is therefore incorrect.  In \S\ref{sec:periods}, we demonstrate how the Z20 method for selecting the orbital period ratios is prone to bias.  In \S\ref{sec:discussion} we consider the peas-in-a-pod pattern from the framework of transit SNRs instead of planet radii.  In our continued examination of the data, we find evidence supporting that the patterns in the \Kepler\ planetary systems are astrophysical in origin and not the result of detection bias.  We conclude in \S\ref{sec:conclusion}.

\section{Peas Examples: High-Multiplicity Systems of Small Planets \label{sec:evidence}}
For any detected planet, a larger planet in its place would have produced a deeper transit and also would have been detectable.  Thus, detection bias cannot explain the systems with multiple small, similarly sized planets.  Consider K03158 \citep[Kepler-444,][]{Dupuy2016} as an example (Figure \ref{fig:peas}).  All five transiting planets are roughly Mars-sized.  An Earth-sized planet could have been detected if any of the Mars-sized planets were in fact Earth-sized.  In the CKS sample, there are 10 systems with four or more transiting planets that are all smaller than 1.8~\rearth:
K03158,
K02169,
K02029,
K01151,
K04032,
K00623,
K00719,
K02722,
K00671, and
K01306 (Figure \ref{fig:peas}).  In these 10 systems, the correlation of a planet's size with the size of its neighbor cannot be explained by detection bias. Furthermore, in many high-multiplicity systems, the planet orbits are sufficiently compact that a point-mass would not be stable between the planets \citep{Fang2013}.

\section{Null Hypothesis Testing\label{sec:mock}}
Even stronger support for an astrophysical origin of the peas-in-a-pod patterns can be established by falsifying a null hypothesis related to each pattern.  In regard to the correlation between neighboring planet sizes (Figure \ref{fig:rp-vs-snr}, top left), W18 posed the following null hypothesis: the size of a planet is random and independent of the size of its neighbor.  We tested whether the null hypothesis, convolved with \Kepler's detection bias, could explain the correlation in the observed planet radii.  

Z20 has called the results of W18 into question and proposed an alternative form of hypothesis testing.  In \S\ref{sec:radii}, we review the logic of the W18 null hypothesis testing method. We then examine the modifications proposed in Z20 in \S\ref{sec:snr}.

\subsection{Resampling Planet Radii \label{sec:radii}}
To test our null hypothesis, we constructed synthetic planetary systems using a bootstrap resampling method.  Our synthetic systems were identical to the observed systems, except we drew the planet sizes at random, with replacement.  The action of drawing the planetary radius at random with replacement immediately produces an instance of the null hypothesis: the radius of each planet is random, with no dependence on the size of its neighbors.

However, as Z20 noted, we do not know the underlying distribution of planet radii, which leads to the question, from which distribution of planet radii should we randomly draw?  In principle, we are free to propose \textit{any} planet radius distribution, so long as it correctly replicates the null hypothesis statement (the radii of adjacent planets are not correlated) \textit{before} we apply detection biases.  In W18, we tried drawing planet sizes from two different distributions: the observed distribution of planet radii, and a log-normal distribution of planet radii (a function weighted toward many more small planets than large planets).  

After populating each synthetic planetary system with new, randomly drawn planet sizes that were independent of the sizes of their neighbors, we applied \Kepler's detection bias to our synethic systems.  We  calculated the signal-to-noise ratio (SNR) based on the following equations:
\begin{equation}
\label{eqn:snr}
    \mathrm{SNR} = \frac{(\rpl/\rstar)^2 \sqrt{3.5 \mathrm{yr}/P}}{\mathrm{CDPP_{6hr}}\sqrt{6 \mathrm{hr} /T_0}}
\end{equation}
where $P$ is the planet orbital period, \rstar\ is the stellar radius, CDPP$_\mathrm{6hr}$ is the combined differential photometric precision, a measure of the photometric variability on a timescale of six hours \citep{Christiansen2012}, and $T_0$ is the transit duration, which depends on the stellar density $\rho_{\star}$ and is given by
\begin{equation}
    T_0 = 13 \mathrm{hr} (P/\mathrm{1yr})^{1/3}(\rhostar/\rho_\odot)^{-1/3}.
    \label{eqn:t0}
\end{equation}
This set of equations ignores several effects that contribute to the SNR at the level of $\sim10\%$, including the impact parameter of the planet, the orbital eccentricity of the planet, the fact that some stars were not observed for exactly 3.5 years, and slight differences between the multiple event statistic (MES, the criterion used for detection by the Kepler team) and SNR, but these other effects are extraordinarily difficult to disentangle.  Because 70\% of the CKS multis have SNR $> 20$, which is signficantly above the detection threshold for most orbital periods in our sample \citep{Christiansen2016}\footnote{for MES $\ge10$ and $P < 40$ days, \Kepler's detection efficiency was 80\%}, inaccuracies of our computed SNR of order 10\% are unimportant within the scope of this analysis.

In W18, we discarded sythetic planets that were too small to detect (SNR $< 10$), thus conservatively mimicking the \Kepler\ detection bias.\footnote{Kepler actually used MES $\ge$ 7.1 as its threshold, but it likely missed a large number of transiting planets with 7.1 $\le$ MES $<$ 10, especially in multi-planet systems \citep{Zink2019}.}  Synthetic planets that were small, at long periods, or orbiting stars that were noisy and/or large were likely to be missed.  

However, as Z20 noted, our method in W18 had some non-ideal attributes.  (1) Drawing from the observed distribution of planet sizes rarely populated stars with planets smaller than 1\,\rearth, and (2) discarding planets can result in a reduction of the number of detected synethtic planets of $\sim20\%$.  Therefore, here, we make two minor changes to our primary analysis in W18:
\begin{enumerate}
    \item We draw planet radii from a log-normal distribution {$\displaystyle \ln(\rpl/\rearth)\in {\mathcal {N}}(\mu ,\sigma ^{2})$} with $\mu=0$ and $\sigma=1$, and
    \item So long as a planet is too small to be detected, we draw a planet size again at random from our distribution, repeating until the drawn planet radius produces SNR $\ge 10$.\footnote{For those who prefer that the undetected planets are discounted rather than redrawn, this strategy was adopted in W18, and the results did not differ significantly from what we present here.}
\end{enumerate}

As stated above, hypothesis testing is valid for any proposed radius distribution, so long as the null hypothesis condition is met, justifying our switch to the log-normal planet radius distribution.  Also, it is reasonable to keep drawing planets until we draw one of sufficient size to be detected because this procedure reproduces our selection criteria\footnote{\citet{Xie2016} used this technique to fully populate synthetic multi-planet systems.}: the CKS multis are systems that were selected for having multiple detected planets, and ensuring that the same planet multiplicities are generated in every trial replicates this selection bias.  
\begin{figure}
    \centering
    \includegraphics[width=0.32\textwidth]{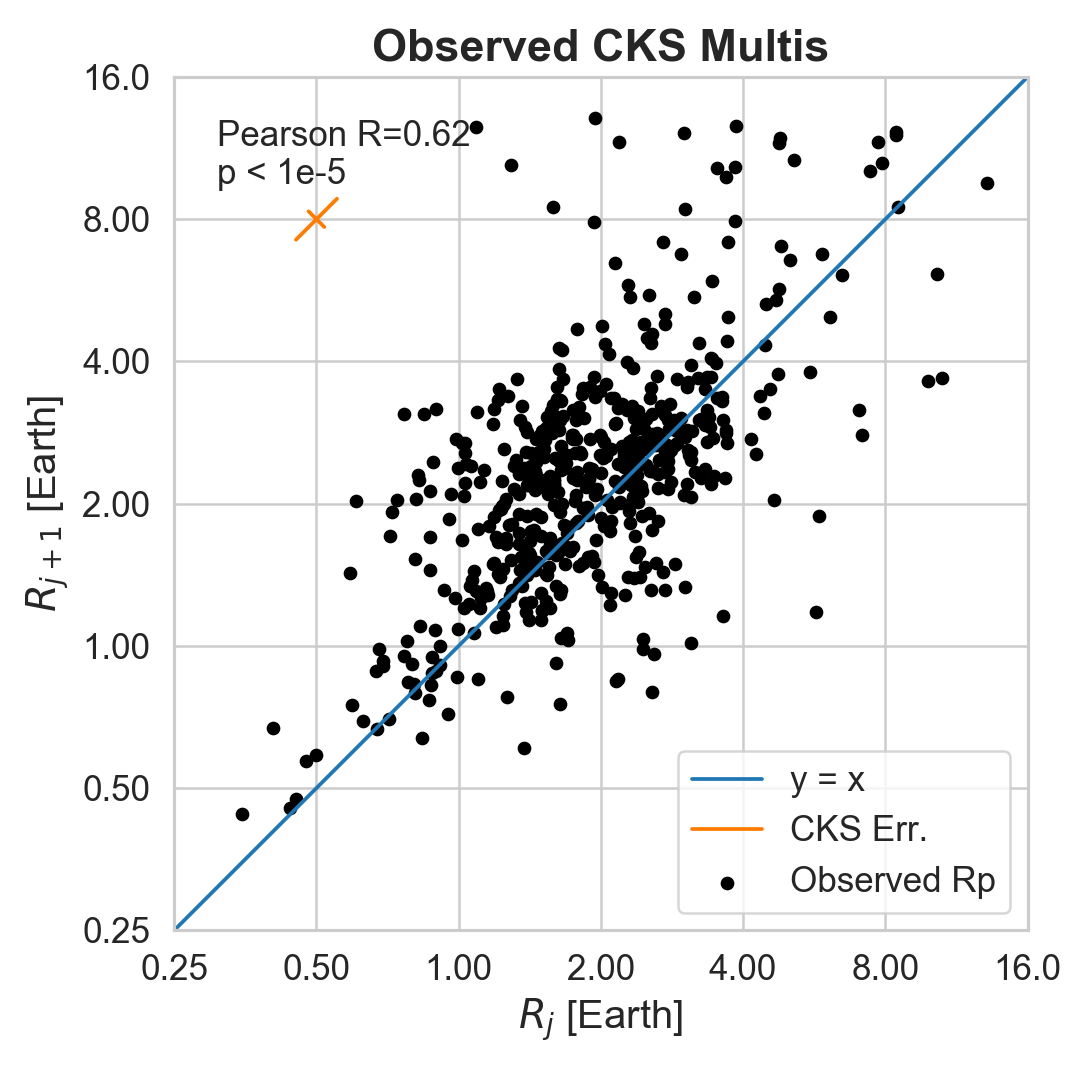}
    \includegraphics[width=0.32\textwidth]{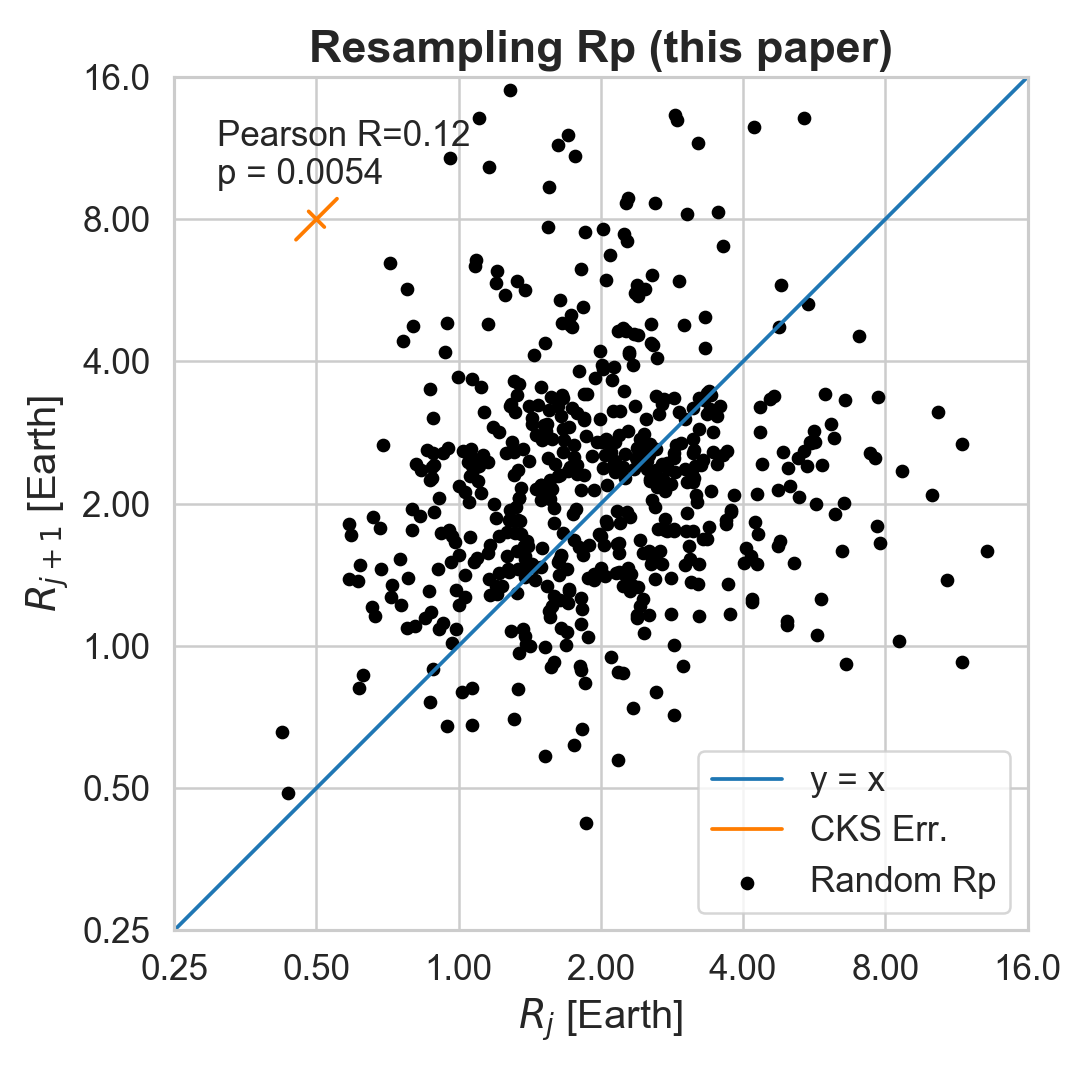}
    \includegraphics[width=0.32\textwidth]{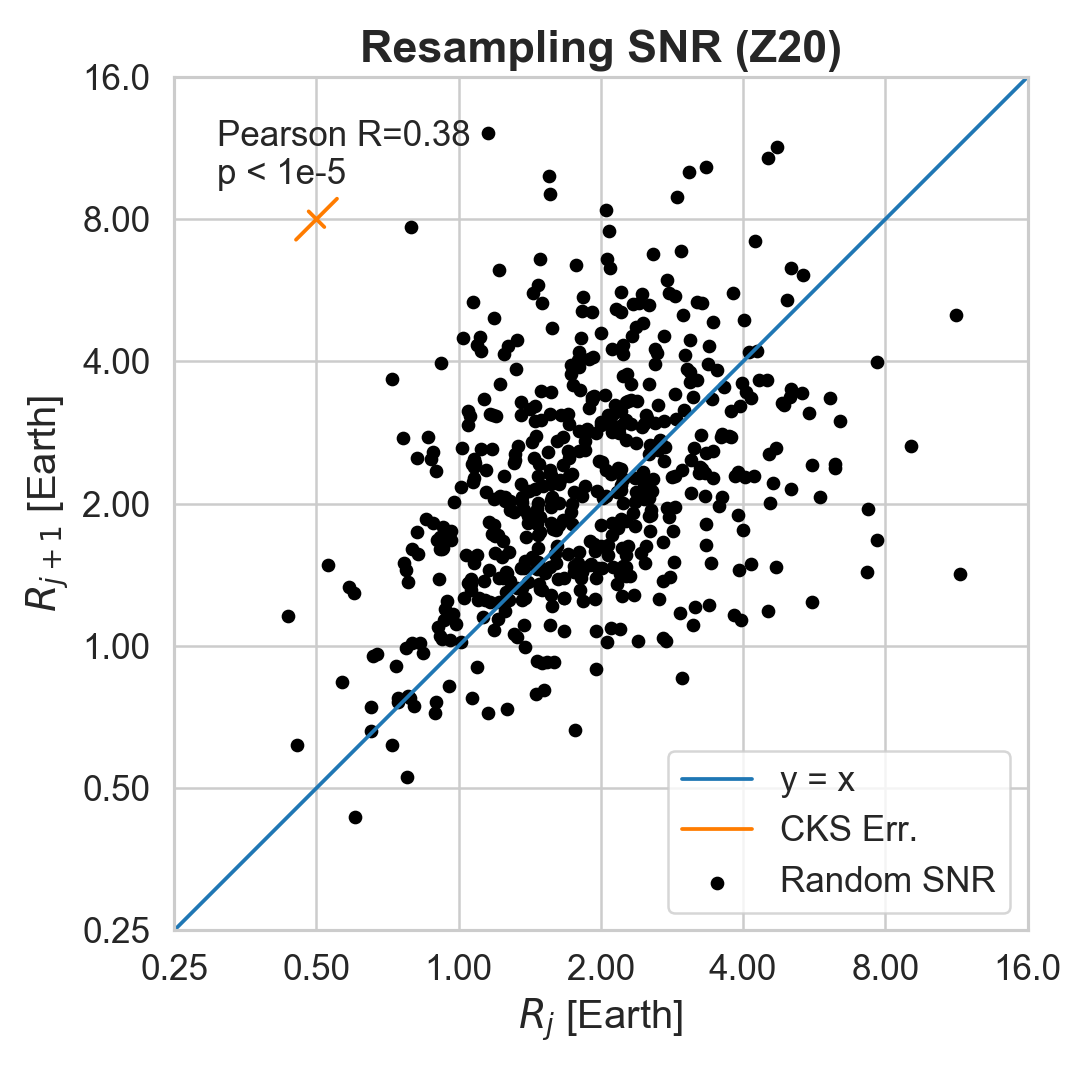}
    \includegraphics[width=0.32\textwidth]{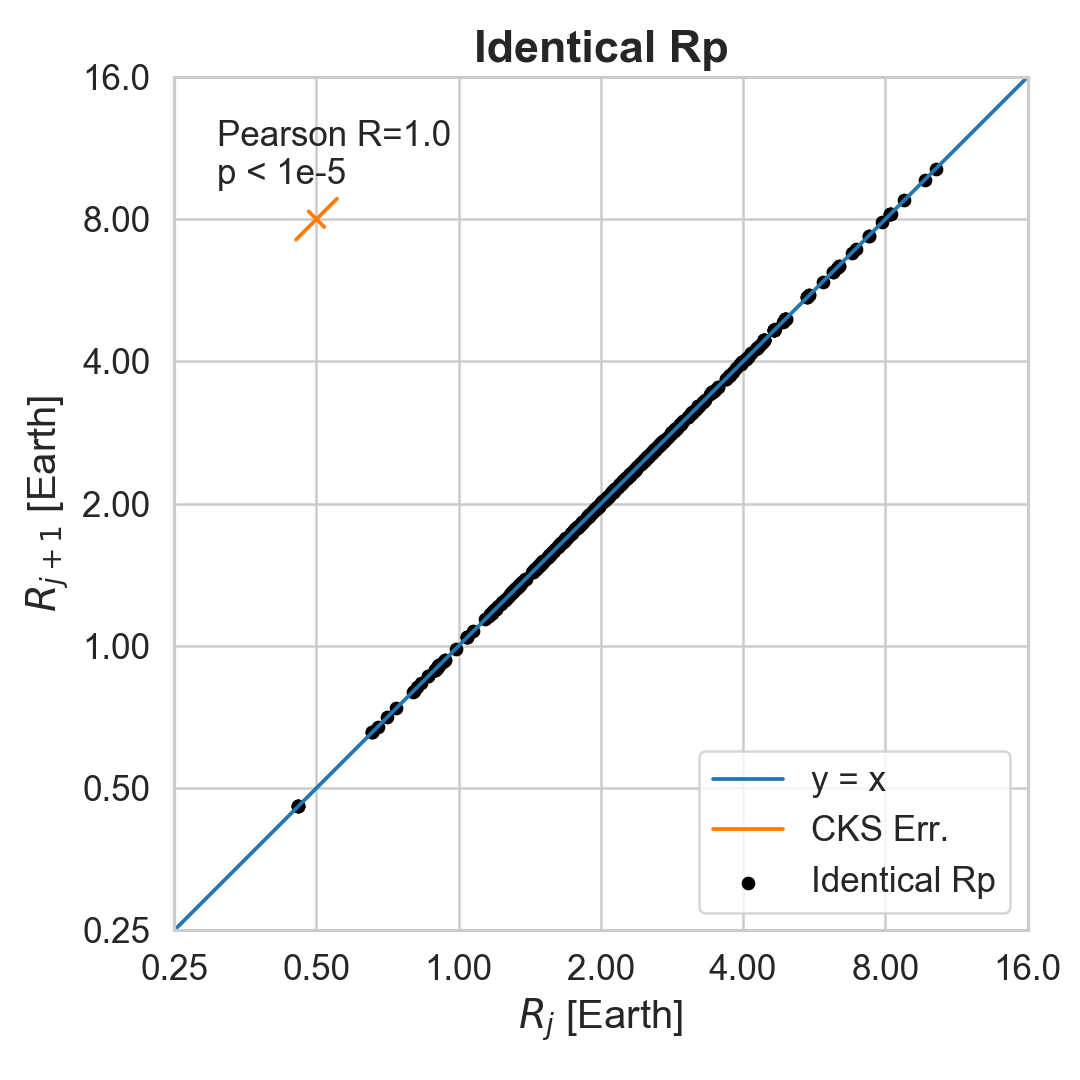}
    \includegraphics[width=0.32\textwidth]{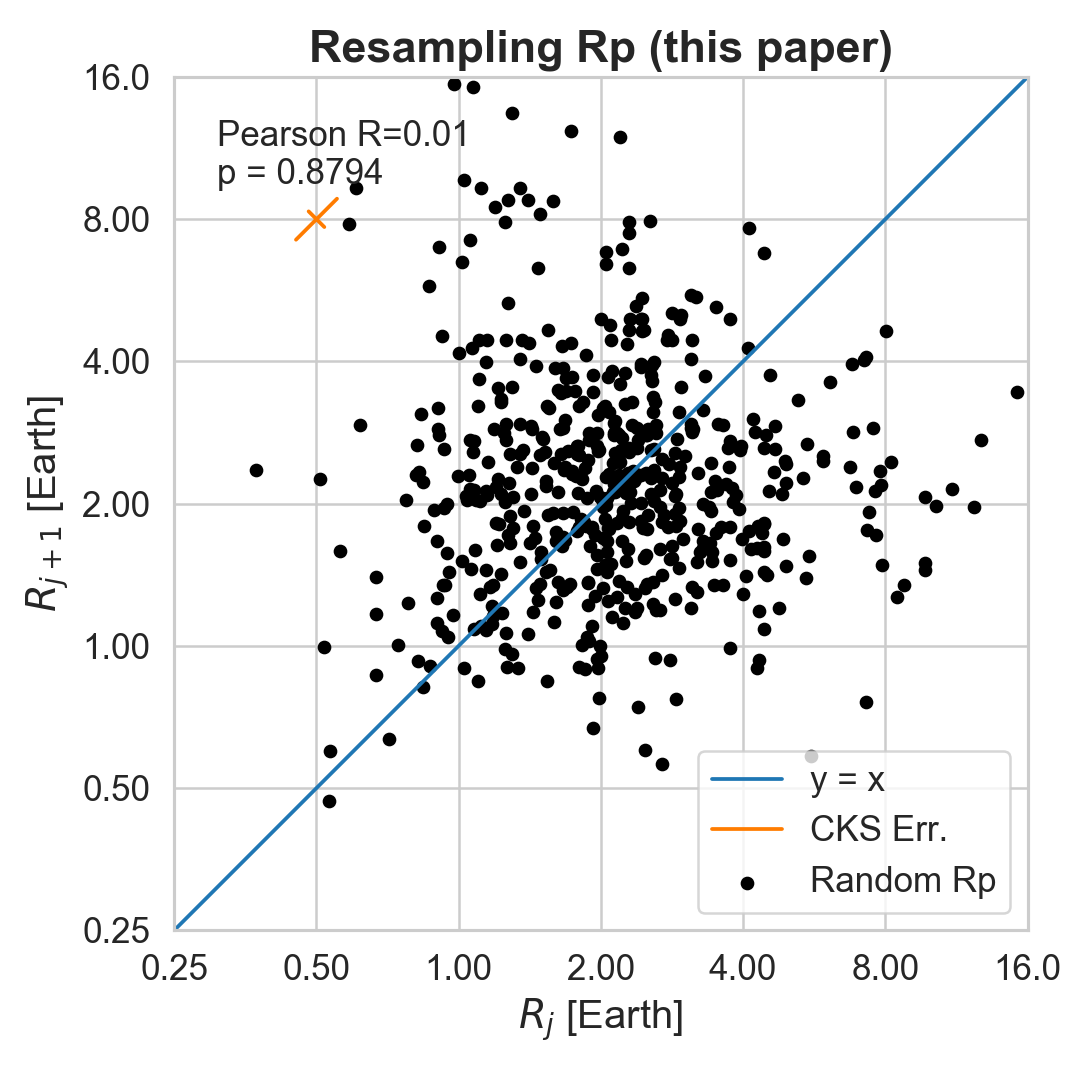}
    \includegraphics[width=0.32\textwidth]{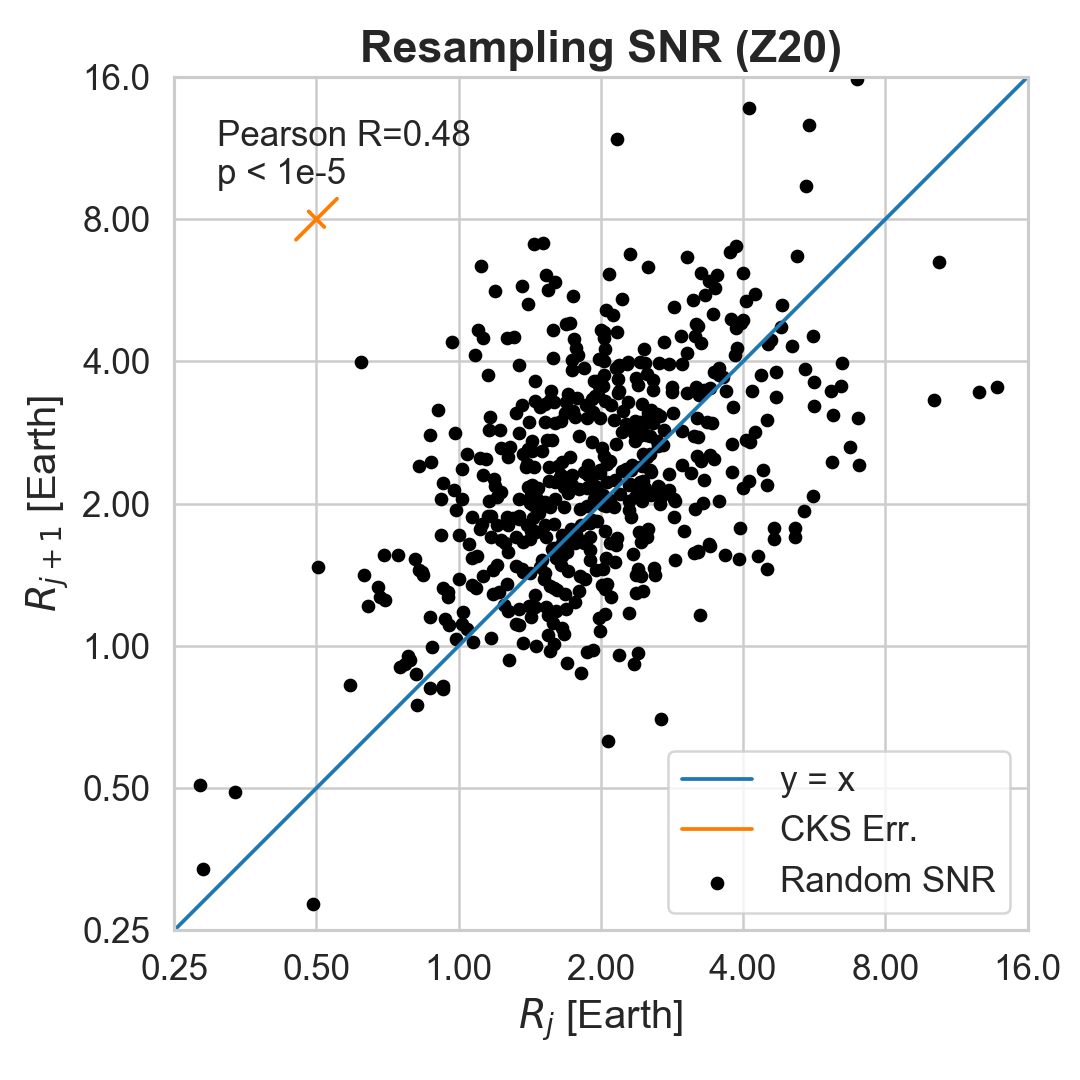}
    \caption{\textbf{Drawing SNR at random produces correlated planet radii in two samples in which the underlying planet radii are correlated.}  Top left: the observed sizes of planets and their adjacent neighbors in the CKS multis sample.  Top middle: resampling the CKS multis sample by drawing planet radius at random and also applying \Kepler's detection biases does not reproduce the observed correlation.  Top right: drawing transit SNR at random results in planet radii that are correlated. Bottom left: the radii of adjacent planets in mock systems in which, by construction, the radius of each planet is identical to its neighbors.  Bottom middle: resampling the radii of the planets by drawing \rpl\ at random and applying \Kepler's detection biases does not produce correlated planet sizes.  Bottom right: resampling the radii of the planets by drawing the transit SNR at random and converting it to \rpl\ produces correlated planet sizes.  \textit{This test demonstrates that the presence of a correlation when resampling the transit SNRs cannot be used as evidence that the underlying planet radii are not correlated.}}
    \label{fig:rp-vs-snr}
\end{figure}

We repeated our resampling algorithm 1000 times for each planetary system.  None of the trials reproduced a correlation between planet sizes with a similar Pearson-R value or significance to what we observed in the distribution of detected transiting planet radii.  One example of a resampling trial is shown in Figure \ref{fig:rp-vs-snr} (top center).  Aggregated over 1000 trials, the Pearson-R values from sampling our null hypothesis were R$=0.023\pm0.044$, whereas the correlation in the observed data was R$=0.62$.  Therefore, with $>10\sigma$ confidence, we rule out \Kepler\ detection bias as the source of the correlated planet radii.

\subsection{Resampling Transit SNR \label{sec:snr}}
Z20 have performed a new kind of test which they interpret as showing that the properties of a given planet are ``independent of the properties of their siblings.''  The test in Z20 is based on drawing a transit signal-to-noise ratio (SNR) at random and then converting to planet radius.  Their reasoning for drawing transit SNR instead of planet radius is that (1) the SNR distribution is ``more fundamental''\footnote{We disagree with this assertion but consider how to treat SNR as the parameter of interest in \S\ref{sec:discussion}.}, and (2) this procedure is a ``shortcut'' for resampling the planet radii without the step of testing whether the newly drawn planet radius would be detected.

Their procedure is as follows: for each planet, a transit SNR is drawn at random from the observed distribution of SNRs.  The planet orbital period and stellar properties are unchanged.  The randomly drawn (new) SNR is used in the computation of a new radius for the planet based on equations \ref{eqn:snr} and \ref{eqn:t0}.  Rearranging for $\rpl$, we have 
\begin{equation}
    \rpl = \rstar(\mathrm{CDPP_{6hr}})^{1/2} (\mathrm{SNR})^{1/2} (3.5 \mathrm{yr}/P)^{-1/4}(6 \mathrm{hr} /T_0)^{1/4}
    \label{eqn:rp-snr} 
\end{equation}
The new planet radius can be computed with the above equation.  However, with some algebraic substitutions, this equation can be rewritten in a much simpler form:
\begin{equation}
\label{eqn:footnote}
    R_{\mathrm{p,new}} =  R_{\mathrm{p,obs}}\sqrt{\mathrm{SNR_{new}}/\mathrm{SNR_{obs}}}
\end{equation}
where the subscript ``obs'' indicates the observed value, and the subscript ``new'' indicates the resampled (synthetic) value (Z20, footnote 7).  

The planet radii produced by drawing the transit SNR at random are strongly correlated, unlike the planet radii produced by drawing \rpl\ at random (Figure \ref{fig:rp-vs-snr}, top right).  

Why do the two methods for drawing new planet radii produce different results?  Of their method, Z20 assert, ``correlated S/N values do not necessarily [produce] correlated [planet] sizes [around the same star].''\footnote{S/N is the same as SNR.}  However, within the CKS data set, drawing random SNR values and applying them to a single star does indeed produce correlated planet radii.

To better understand the difference between the W18 and Z20 resampling methods, we construct a mock universe in which all of the planet radii are identical to their neighbors (Figure \ref{fig:rp-vs-snr}, bottom left).  Drawing the planet radius at random (as in W18) produces systems in which the planet radius is uncorrelated with the size of its neighbor (bottom center). However, drawing the SNR at random (as in Z20) produces synthetic systems in which neighboring planet radii are still correlated (bottom right).  The correlation produced by resampling SNR occurs even when there is an underlying planet size correlation.  Therefore, performing the Z20 test and seeing a correlation in the sizes of adjacent planet radii is insufficient to rule out the hypothesis that the underlying planet radii are correlated.

\begin{figure}
    \centering
    \includegraphics[width=0.32\textwidth]{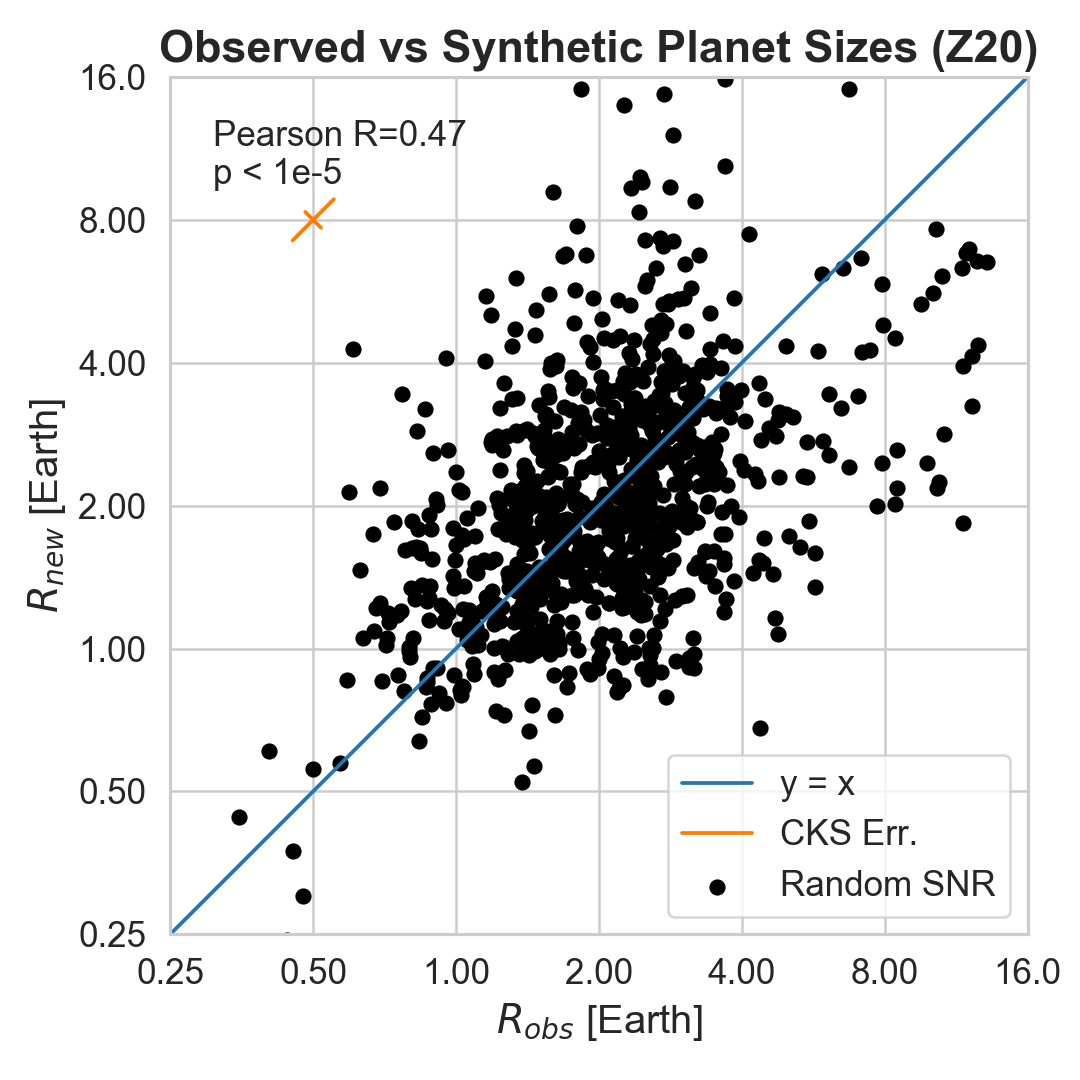}
    \caption{\textbf{Drawing SNR at random produces non-random, non-independent planet radii.}  The x-axis is the observed radius of each planet $R_{\mathrm{obs}}$, and y-axis is the new (resampled) planet radius $R_{\mathrm{new}}$ computed by drawing SNR at random and converting to \rpl\ via equation \ref{eqn:rp-snr}.  There is a strong correlation (Pearson-R=0.42) between $R_{new}$ and $R_{\mathrm{obs}}$, which it is replacing.  Therefore, the synthetic planetary systems constructed by drawing SNR at random have a strong resemblance to the underlying planetary system architecture.  The non-randomness of the planet radii accounts for the eventual correlation between the sizes of adjacent planets in the Z20 resampled systems (Figure \ref{fig:rp-vs-snr}, right panels).}
    \label{fig:rp_synth}
\end{figure}

What produces the correlated planet radii when the transit SNR is drawn at random?  Equation \ref{eqn:footnote} reveals that $R_\mathrm{p,new}$ is proportional to $R_\mathrm{p,obs}$, meaning that the newly computed planet radius is correlated with the observed planet radius.  We show this in Figure \ref{fig:rp_synth}. When we draw a new SNR at random and convert it to a planet radius via equation \ref{eqn:rp-snr}, the new planet radius is correlated with the old planet radius, as we would expect from equation \ref{eqn:footnote}.  Thus, if the observed (and underlying) planet radii are correlated, then this correlation will be partially preserved.  The multiplication by $\sqrt{\mathrm{SNR_{new}}/\mathrm{SNR_{obs}}}$ will introduce some random variation, but not enough to erase the information about the underlying planet size correlation, since the dynamic range in $\sqrt{\mathrm{SNR_{new}}/\mathrm{SNR_{obs}}}$ is comparable to the the dynamic range of $R_\mathrm{p,obs}$.  This is why in the right panels of Figure \ref{fig:rp-vs-snr}, the planet radii that are computed by drawing SNR at random are correlated, but with a lower Pearson-R value than the original planet radii (left panels).

In summary, we can explain the correlated planet radii produced in the bootstrap resampling method of Z20 as coming from a failed assumption.  Z20 assume that drawing SNR at random should also produce random planet radii.  However, this is not the case;  each synthetic planet radius produced by the Z20 method is strongly correlated with the observed planet radius, and so \textit{the Z20 resampled planet radii are not random.}  When the underlying planet radii are correlated, the Z20 resampled planet radii are also correlated.  This aspect of the Z20 method makes it unsuitable for testing the null hypohtesis: their algorithm is not able to distinguish whether the underlying planet radii are correlated or whether detection bias produces the observed correlation.

Even if the Z20 hypothesis testing method were not flawed, note that it does not reproduce the strength of the correlation in the data.  In Figure \ref{fig:rp-vs-snr}, drawing SNR at random produces neighboring planet radii that are correlated with Pearson-R=0.45, even though the observed population of neighboring planet radii were more strongly correlated (Pearson-R=0.62).

\section{On Period Ratio Sampling} \label{sec:periods}
In W18, we found that the orbital spacing between pairs of planets in the same \Kepler\ system is regular.  More specifically, the period ratio between one pair of adjacent planets, $\mathcal{P}_j$, and the next pair of adjacent planets, $\mathcal{P}_{j+1}$ is correlated for systems that have at least two pairs of planets with $\mathcal{P} < 4$.  Z20 noted that we did not justify our choice of $\mathcal{P} < 4$ in W18, and they chose an alternative cut.  

In the context of the limitations of \Kepler's sensitivity, it is possible to detect triples of planets so long as the product of their period ratios is less than some constant (Z20 chose 25; $\mathcal{P}_j \mathcal{P}_{j+1} < 25$).  The problem with the Z20 cut is that it selects triples where the period ratio of one pair of planets is dependent on the period ratio of the other pair of planets, and this dependency creates bias.  Consider a hypothetical three-planet system with an innermost planet at $P = 2$ days, the second planet at $P = 40$ days (making the first period ratio $\mathcal{P}_j = 20$), and the third planet at $P = 50$ days (making the second period ratio $\mathcal{P}_{j+1} = 1.25$).  The product of the period ratios is 25, making this system just within the Z20 threshold.  However, if we change the orbital period of the third planet to be $P = 60$ days, the product exceeds the threshold and so this triple would not be counted.  This is problematic because the value we drew for $\mathcal{P}_j$ dictated that the value for $\mathcal{P}_{j+1}$ had to be within a certain range ($<1.25$).  In other words $\mathcal{P}_{j+1}$ is not independent of $\mathcal{P}_j$.  

\begin{figure}
    \centering
    \includegraphics[width=\textwidth]{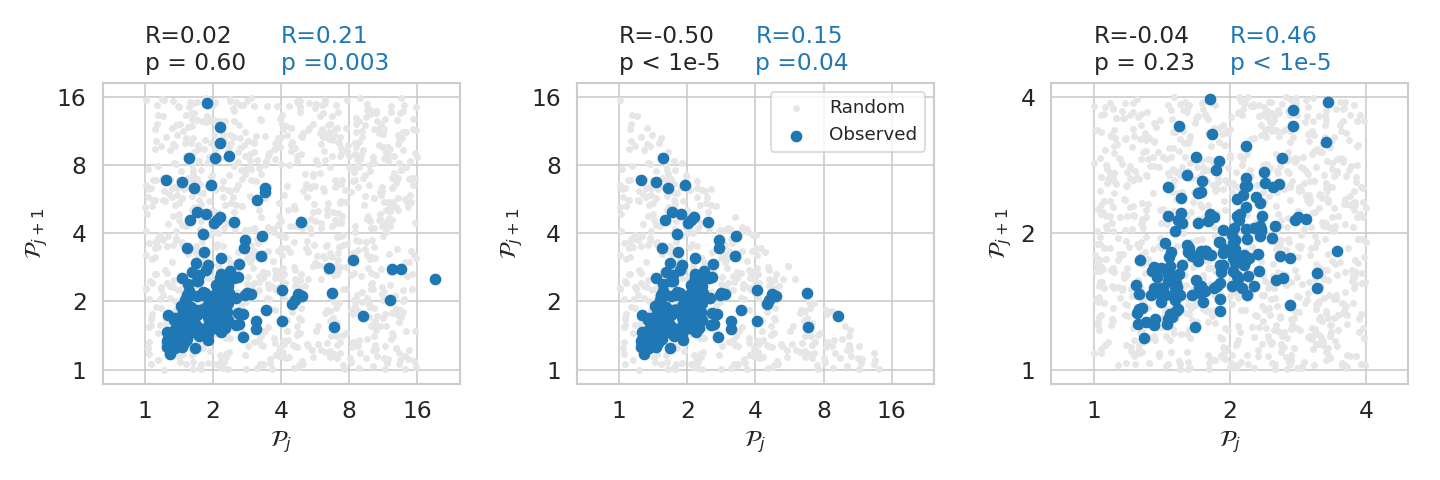}
    \caption{A comparison of the Pearson-R correlations of the observed CKS period ratios (blue) and draws from a random uniform distribution (gray), after applying various cuts. Left:  the gray points are drawn randomly from $0 < \mathrm{log_2}(\mathcal{P}) < 4$.  There is no correlation between the randomly drawn points.  Center: same as left, but after applying rejection sampling $\mathrm{log_2}(\mathcal{P}_j) + \mathrm{log_2}(\mathcal{P}_{j+1}) < 4$ (i.e., the product of the period ratios is $<16$).  The rejection sampling induces a strong negative correlation (Pearson-R=-0.5, $p < 10^{-5}$) in the randomly drawn period ratios (gray).  Note that the CKS distribution (blue, R=0.15, p=0.04) is inconsistent with a random distribution of period ratios.  Right: the need to keep $\mathrm{log_2}(\mathcal{P}_j)$ and $\mathrm{log_2}(\mathcal{P}_{j+1})$ independent motivates our choice of $\mathcal{P}_j < 4$ and $\mathcal{P}_{j+1} < 4$.  Within this regime, there is a strong correlation between the period ratios of adjacent planets.}
    \label{fig:diagonal}
\end{figure}

In general, a mathematical dependence between two variables can induce a correlation.  To test the correlation induced by a dependence between $\mathcal{P}_{j+1}$ and $\mathcal{P}_j$, we compare the CKS period ratio distribution to a distribution of randomly drawn period ratios.  To produce our random set of period ratios, we draw pairs of points ($x$, $y$) from a uniform distribution on $((0,4),(0,4))$, where $x = \mathrm{log}_2\mathcal{P}_j$, $y = log_2\mathcal{P}_{j+1}$, and $x$ and $y$ are drawn independently (Figure \ref{fig:diagonal}, left panel, black points).  We demonstrate with a Pearson-R test that there is no underlying correlation between $x$ and $y$ when drawn this way.  However, if we down-select our draws of $(x,y)$ to only include pairs that meet $x + y < 4$ (the product of the period ratios is $<16$), mimicking a detection bias, we introduce a significant negative Pearson-R correlation: R$=-0.50$, $p < 10^{-5}$ (center panel).  The Pearson-R correlation of the observed period ratios satisfying this cut (shown in blue points) is R$=0.15$, $p < 10^{-5}$; this is not a strong correlation, but it is significantly different from the strong anti-correlation we would expect if the period ratios were indeed random.  This evidence strongly disfavors the Z20 conclusion that the period ratios of the planets are random.

If we further down-select our data to $x, y < 2$ ($\mathcal{P} <4$ and $\mathcal{P}_{j+1} <4$), thereby excluding the regions where $x$ and $y$ were dependent, we indeed see a strong Pearson-R correlation in the CKS data, and no correlation in the randomly drawn $x$ and $y$ values (Figure \ref{fig:diagonal} right panel).  In compact planetary systems, the period ratios are indeed regular.  The regularity (or lack thereof) for larger period ratios
has yet to be tested.

\section{Discussion\label{sec:discussion}}
We have shown that the method of drawing SNR and converting to planet radius, as done in Z20, does not sample planet radii at random.  This makes the method unsuitable for testing any null hypothesis in which it is necessary to produce random, independent planet radii.  Like W18, \citet{Millholland2017} and \citet{Ciardi2013} demonstrate patterns in planet radii by falsifying hypotheses in which the planet radii are random.  The Z20 method is therefore unsuitable for testing the null hypotheses in those papers as well.

Z20 claims that the transit SNR is a better parameter to examine than the planet radius.  Even if the transit SNR were chosen as the parameter of interest, the questions about peas in a pod would need to be reframed around this parameter.  We would ask whether adjacent transiting planets have correlated transit SNRs (they do, with Pearson-R=0.37, $p<10^{-5}$), and whether a bootstrap resampling algorithm in which we draw the SNRs of the transiting planets at random could reproduce the observed, correlated transit SNRs (it cannot; in 1000 trials, the mean and standard deviation of the Pearson-R value was $0.0\pm0.044$, meaning that the correlation between adjacent transiting planets' SNRs is significant with 8-$\sigma$ confidence).

\section{Conclusion \label{sec:conclusion}}
We have demonstrated that the statistical tests performed by Z20 were not suitable to test the role of detection bias in the peas-in-a-pod pattern.  We summarize the main flaws below. (1) When the underlying planet radii are correlated, drawing SNR at random and converting to a planet radius results in non-random planet radii, and so this method is not suitable for null hypothesis testing. (2) The CKS distribution of period ratios is inconsistent with random period ratios, and the period ratios of adjacent planets are indeed correlated when the period ratio is less than 4.  Because of these methodological flaws, the conclusions of Z20 regarding the peas in a pod pattern, or other patterns in which it is necessary to sample planet radii at random, are not applicable. 

Furthermore, the analysis of Z20 overlooked evidence in favor of an astrophysical interpretation of the peas-in-a-pod pattern.  Namely, there are many systems in which multiple similarly-sized small planets were detected, yet large planets were not detected.

\acknowledgments
{The authors thank Sarah Millholland, David Ciardi, Greg Laughlin, Jack Lissauer, Songhu Wang, Geoff Marcy, and the anonymous referees for their comments on various versions of this manuscript.  We would also like to acknowledge the significant role of the many scientists who gathered, curated, and presented data from both the NASA \Kepler\ spacecraft and the W.M. Keck Observatory to create the California Kepler Survey catalog.
LMW acknowledges support from the Beatrice Watson Parrent Fellowship.}

\bibliography{main}{}
\bibliographystyle{aasjournal}

\end{document}